\documentclass{conm-p-l}     

\usepackage{amsmath}     
\usepackage{amssymb}     
\usepackage{amsfonts}    
\usepackage{amsthm}      

\theoremstyle{plain}
\newtheorem{thm}{Theorem}[section]
\newtheorem{lem}[thm]{Lemma}
\newtheorem{prop}[thm]{Proposition}
\newtheorem{cor}[thm]{Corollary}

\theoremstyle{definition}

\newtheorem{defn}[thm]{Definition}
\theoremstyle{remark}

\newcommand{\thereals}{\ensuremath{\mathbb{R}}}

\newcommand{\manifold}[1]{\ensuremath{\mathcal{#1}}}

\newcommand{\spacetime}[2]{\ensuremath{(\manifold{#1},#2)}}
\newcommand{\curvefamily}[1]{\ensuremath{\mathcal{#1}}}

\newcommand{\goesto}{\ensuremath{\rightarrow}}

\renewcommand{\mapsto}{\rightarrow} 
 
\newcommand{\D}{\displaystyle}

\newcommand{\abndry}[1]{\ensuremath{\mathcal{B}(\manifold{#1})}} 
\newcommand{\envbndry}[2]{\ensuremath{\partial_{#1}\manifold{#2}}}

\renewcommand{\vec}[1]{\ensuremath{#1}}
\newcommand{\ten}[1]{\ensuremath{#1}}

\normalsize

\begin{document}
\title[Curvature in abstract boundary singularity theorems]{Curvature singularities and abstract boundary singularity theorems for space-time}
\author{Michael J. S. L. Ashley}
\address{Department of Physics \\ Faculty of Science \\ The Australian National University \\ Canberra ACT   0200 \\ AUSTRALIA}
\curraddr{Center for Gravitational Wave Physics \\ Department of Physics \\ Eberly College of Science \\ The Pennsylvania State University \\ 104 Davey Laboratory - PMB 005 \\ University Park, PA 16802-6300}
\email{ashley@gravity.psu.edu}
\author{Susan M. Scott}
\address{Department of Physics \\ Faculty of Science \\ The Australian National University \\ Canberra ACT   0200 \\ AUSTRALIA}
\email{Susan.Scott@anu.edu.au}

\thanks{This research was supported by an Australian Postgraduate Award (MJSLA) and in part by a Fulbright Postgraduate Award (MJSLA)}

\begin{abstract}
The abstract boundary construction of Scott and Szekeres is a general and flexible way to define singularities in General Relativity.
The abstract boundary construction also proves of great utility when applied to questions about more general boundary features of space-time.
Within this construction an essential singularity is a non-regular boundary point which is accessible by a curve of interest (e.g.\ a geodesic) within finite (affine) parameter distance and is not removable.
Ashley and Scott proved the first theorem linking abstract boundary essential singularities with the notion of causal geodesic incompleteness for strongly causal, maximally extended space-times.
The relationship between this result and the classical singularity theorems of Penrose and Hawking has enabled us to obtain abstract boundary singularity theorems.
This paper describes essential singularity results for maximally extended space-times and presents our recent efforts to establish a relationship between the strong curvature singularities of Tipler and Kr\'olak and abstract boundary essential singularities.
\end{abstract}
\subjclass[2000]{Primary: 53C50, 83C75; Secondary: 53C23, 57R15}
\keywords{singularity, abstract boundary, space-time}

\maketitle

\section{Introduction}
There are many convincing reasons for using the abstract boundary construction ($a$-boundary) of Scott and Szekeres \cite{Scott/Szekeres:1994} when considering the issue of singularities for space-time as opposed to using the $g$, $c$ or $b$-boundary constructions of Geroch \cite{Geroch:1968a}, Geroch, Kronheimer and Penrose \cite{Geroch/Kronheimer/Penrose:1972}, and Schmidt \cite{Schmidt:1971}, respectively.
The $g$, $c$ and $b$-boundaries are plagued with problems of a topological nature, and in certain circumstances produce non-intuitive results \cite{Bosshard:1979, Szabados:1988}.
An overview of the literature dealing with this issue has been given by Ashley \cite[Chapter~2]{Ashley:phdthesis}.
The common theme presented there is that the $a$-boundary provides a superior alternate structure in which to consider space-time singularities and other boundary features of space-time.
Recently it has become apparent that the $a$-boundary formalism might allow the proof of curvature singularity theorems for space-time.
This is a result of direct application of the abstract boundary classification scheme to Theorem 3 of Hawking \cite[p.~271]{Hawking/Ellis:1973:book}\footnote{It may be beneficial to the reader to compare this proof to that given in \cite{Hawking:1967}.}.
The recent result of Ashley and Scott \cite[Theorem~4.12]{Ashley:phdthesis} applies directly to the maximally extended space-times covered by Hawking's theorem and the link this forms with curvature appears exploitable to produce a curvature singularity theorem.

In this paper we briefly describe the present state of research into singularity theorems for space-time and our recent attempts to include curvature in this framework.
We then formulate what we consider to be the absolute minimum considerations for the proof of a curvature singularity theorem applicable to physically realistic situations.
Finally, we review how much of this structure is already in place and state which additional results are required to complete any such theorem.

\section{Singularity Theorems for Space-time}
The earliest attempts to produce theorems about the existence of singularities in space-time were negative in nature.
It was believed that singularities in the then known cosmological and stellar model space-times were due to their highly symmetrical nature.
The view that singularities would not be a feature of general solutions was put forward by Lifshitz and Khalatnikov  \cite{Lifshitz/Khalatnikov:1963} where it was proposed that Cauchy data giving rise to singularities comprised a set of measure zero in the set of all possible Cauchy data. 
Those authors later retracted this view in \cite{Belinskii/Khalatnikov/Lifshitz:1970}.

The first positive theorems about the existence of singularities were pioneered by Penrose \cite{Penrose:1965c} and Hawking \cite{Hawking:1967}.
Instead of dealing with the properties of solutions of the Einstein Field Equations directly, the Penrose-Hawking singularity theorems applied the techniques of differential topology and used only the most fundamental topological properties of the space-time and the behaviour predicted by General Relativity.
In these singularity theorems timelike, null or causal geodesic incompleteness is deduced from a structure which is roughly composed of three parts: 
\begin{enumerate}
\item {\bf Energy condition} --- usually the strong energy or null convergence condition,
\item {\bf Causality condition} --- the strong causality or chronology condition,
\item {\bf Closure or Curve Trapping condition} --- these sometimes take the form of a topological condition (e.g.\ the presence of a compact spacelike hypersurface) or a trapping condition on the causal geodesics defined using a differential geometric construct such as the null second fundamental form or expansion scalar.
\end{enumerate}
The energy condition guarantees that gravity is attractive for all matter and energy in the space-time.
The causality condition prevents the escape of causal geodesics from the process of being focussed by curvature through causal violations.
Finally, the trapping condition sets up the elements necessary for gravitational collapse or for some sort of cosmological or primordial singularity.

The primary reason for clarifying the status of these `singularity theorems' is that, as mentioned above, there are two inequivalent philosophies for the definition of singular behaviour for space-time.
Examples of space-time exist (see \cite{Misner:1967}) where geodesic incompleteness occurs without the presence of any curvature pathology.
These situations typically involve \emph{imprisoned incompleteness} \cite[\S~8.5, p.~289]{Hawking/Ellis:1973:book} or \emph{quasi-regularity} \cite{Clarke/Schmidt:1977,Ellis/Schmidt:1977,Ellis/Schmidt:1979}.
So in fact the term `singularity theorem' is a misnomer since these theorems are actually geodesic incompleteness theorems.
It is commonly believed that real singularities must be a type of curvature singularity and present astrophysical evidence supports the formation of singularities only in space-times derived from realistic collapse scenarios.
In the case of primordial singularities, these might be formed from reasonable boundary assumptions (such as those arising from the Weyl curvature hypothesis of Penrose).
Nevertheless, the Penrose-Hawking geodesic incompleteness theorems have had enormous success in most physical situations.
There is, however, strong physical motivation for the specific prediction of `unbounded curvature'.

\section{Problems in Proving Curvature Singularity Theorems}
Much of the difficulty in proving a curvature result has been due to 
\begin{enumerate}
\item the challenge in ascribing boundary points to space-time --- the \emph{boundary construction problem}, and 

\item  the variety of ways in which curvature may not be smooth in a given space-time --- the \emph{curvature pathology classification problem}.

\end{enumerate}
Hence the most significant hurdles to date have not been the proof of a curvature singularity theorem itself but instead in the precise formulation of the theorem which one wants to prove.

The abstract boundary is a promising way to overcome (1) for the following reasons.
As discussed previously, one encounters problems of a topological nature when attempting to generate boundary points for a space-time purely from the pseudo-Riemannian manifold which is used to represent it.
Within this philosophy of boundary construction, it appears that there is no single construction which has all the requisite properties and remains well-behaved.
In essence, the expectation that there exists a single boundary construction, which is physically significant, topologically well-behaved, and easy to construct for any space-time is overly optimistic.
The $a$-boundary, however, uses a completely different philosophy based only on the invariant properties of boundary points in the countless many possible representations of the boundary.
Thus, it avoids the problems described in \cite{Ashley:phdthesis} by not enforcing a particular boundary construction upon a space-time, but instead allowing us to derive invariant results from whichever representation of the boundary is best suited for a specific problem.
In addition, we obtain a precise definition for an essential singularity, which is independent of its envelopment, and for which the previous objections described in \cite{Ashley:phdthesis} do not apply.
We refer the reader to Definitions 37-42 of \cite{Scott/Szekeres:1994} which define $a$-boundary essential singularities.
The reader should also note that we have chosen the family \curvefamily{C} of curves satisfying the bounded parameter property to be the family of affinely parametrised causal geodesics and we employ these curves to define the essential singularities which are central to the following analysis.

Although we have a solution to the boundary construction problem (1), we are still left with the problems provided by (2).
The literature embodies a large number of different definitions of a curvature singularity.
If we consider singularities, not in the sense of curve incompleteness, but as a `place' resisting metric extension, then  we find many specific definitions.
Clarke \cite{Clarke:1975} and Ellis and Schmidt \cite{Ellis/Schmidt:1977} developed schemes aimed at classifying this notion of singularities.
Their work generalised the fundamental definitions of \emph{scalar polynomial curvature singularities} and \emph{parallelly propagated curvature singularities} discussed in Hawking and Ellis \cite{Hawking/Ellis:1973:book} to include not only divergent behaviour but also non-smoothness in the derivatives of the respective curvature based objects.
In addition, Ellis and Schmidt \cite{Ellis/Schmidt:1977} defined these singularities as singular boundary points of the $b$-completion of the space-time.
The previous definitions of Hawking and Ellis \cite{Hawking/Ellis:1973:book} only discussed curves \emph{corresponding} to a curvature singularity and not the boundary points themselves, since no particular boundary construction was employed.
Finally Clarke, 
Ellis and Schmidt 
defined an additional singularity type --- the \emph{quasi-regular singularity} --- for which no type of curvature pathology occurs at all.

In order to gain some feeling for the issue in the abstract boundary context we redefine below and broaden Clarke, 
 Ellis and Schmidt's 
definitions in terms of $a$-boundary concepts.

\begin{defn}[$C^k$ Curvature Singularity]
Let $q \in \envbndry{\phi}{M}\subset\manifold{\widehat{M}}$ be a $C^k$ essential singularity, $k \geqslant 0$.
The boundary point $q$ is termed a \emph{$C^k$ curvature singularity}\footnote{This concept is traditionally referred to in the literature as a \emph{parallelly propagated curvature singularity}.
Similarly, the concept in Definition \ref{scsingularity.defn} is referred to as a \emph{scalar polynomial curvature singularity}.} if there exists a curve $\gamma:[0,a) \mapsto \manifold{M},\: a \in {\thereals}^+$ in \curvefamily{C} which has $q$ as a limit point such that some component $R_{abcd}$ of the Riemann curvature tensor, when evaluated in a parallelly propagated orthonormal tetrad $\{ E_a (s) \}$ along $\gamma(s)$, satisfies the condition 
\begin{equation*}
\limsup_{s \goesto a} \left|R_{abcd;e_1 \ldots e_k}(\gamma(s)) \right| = \infty\;.
\end{equation*}
In this case $\gamma$ is said to \emph{correspond} to the $C^k$ curvature singularity.
\end{defn}

\begin{defn}[$C^k$ Quasi-regular Singularity]
Let  $q \in \envbndry{\phi}{M}\subset\manifold{\widehat{M}}$ be a $C^k$ essential singularity, $k \geqslant 0$.
The boundary point $q$ is termed a \emph{$C^k$ quasi-regular singularity} if it is not a $C^k$ curvature singularity.
\end{defn}

\begin{defn}[$C^k$ Scalar Curvature Singularity]
\label{scsingularity.defn}
Let $q \in \envbndry{\phi}{M}\subset\manifold{\widehat{M}}$ be a $C^k$ curvature singularity.
The boundary point $q$ is termed a \emph{$C^k$ scalar curvature singularity} if there exists a curve $\gamma:[0,a) \mapsto \manifold{M},\: a \in {\thereals}^+$ in \curvefamily{C} which has $q$ as a limit point and a scalar polynomial function $P(x)$ on \manifold{M} composed only of $g_{ab}$, $\eta_{abcd}$ (volume form) and $R_{abcd}$ such that
\begin{equation*}
\limsup_{s \goesto a} \left|P(\gamma(s)) \right| = \infty\;.
\end{equation*}
As above, $\gamma$ is said to  \emph{correspond} to the $C^k$ scalar curvature singularity.
\end{defn}

\begin{defn}[$C^k$ Non-scalar Curvature Singularity]
Let $q \in \envbndry{\phi}{M}\subset\manifold{\widehat{M}}$ be a $C^k$ curvature singularity.
The boundary point $q$ is termed a \emph{$C^k$ non-scalar curvature singularity} if it is not a $C^k$ scalar curvature singularity.
\end{defn}

Using the four definitions above we might then wish to define an abstract boundary point $[p] \in \abndry{M}$ as being one of the above types of singularity --- e.g.\ a curvature singularity --- if it possesses a boundary point representative which is itself that type of singularity. In order to show that this concept is well-defined, the next step in the $a$-boundary approach is then to prove the invariance of this classification under re-envelopment so that the property passes to all boundary point members of the equivalence class.

\begin{prop}
The boundary point $q \in \envbndry{\phi}{M}\subset\manifold{\widehat{M}}$ and the boundary point $p \in \envbndry{\phi'}{M}\subset\manifold{\widehat{M}}'$ are assumed to be equivalent i.e.\ $q \sim p$. If the boundary point $q$ is a $C^k$ curvature singularity, $C^k$ quasi-regular singularity, $C^k$ scalar curvature singularity or $C^k$ non-scalar curvature singularity, then the boundary point $p$ is a singularity of the same type.
\end{prop}

\begin{proof}
If the boundary point $q$ is any of these four types of singularity, then it is, by definition, a $C^k$ essential singularity. It has been shown in Theorem 46 of \cite{Scott/Szekeres:1994} that the property of being an essential singularity passes to the abstract boundary $\abndry{M}$. Thus the boundary point $p$ is also a $C^k$ essential singularity.

Now suppose that $q$ is a $C^k$ curvature singularity. By Definition 3.1, there exists a curve $\gamma:[0,a) \mapsto \manifold{M},\: a \in {\thereals}^+$ in \curvefamily{C} which has $q$ as a limit point such that some component $R_{abcd}$ of the Riemann curvature tensor, when evaluated in a parallelly propagated orthonormal tetrad $\{ E_a (s) \}$ along $\gamma(s)$, satisfies the condition 
\begin{equation*}
\limsup_{s \goesto a} \left|R_{abcd;e_1 \ldots e_k}(\gamma(s)) \right| = \infty\;.
\end{equation*}
Since $p$ covers $q$, then by Theorem 17 of \cite{Scott/Szekeres:1994}, $p$ is also a limit point of the curve $\gamma$ and thereby satisfies the requirements to be a $C^k$ curvature singularity.

If $q$ is a $C^k$ quasi-regular singularity, then by Definition 3.2, it is not a $C^k$ curvature singularity. By the previous argument $p$ cannot then be a $C^k$ curvature singularity and is therefore a $C^k$ quasi-regular singularity.

Now suppose that $q$ is a $C^k$ scalar curvature singularity. This means that $q$ is a $C^k$ curvature singularity and thus $p$ is also a $C^k$ curvature singularity. By Definition 3.3, there exists a curve $\gamma:[0,a) \mapsto \manifold{M},\: a \in {\thereals}^+$ in \curvefamily{C} which has $q$ as a limit point and a scalar polynomial function $P(x)$ on \manifold{M} composed only of $g_{ab}$, $\eta_{abcd}$ (volume form) and $R_{abcd}$ such that
\begin{equation*}
\limsup_{s \goesto a} \left|P(\gamma(s)) \right| = \infty\;.
\end{equation*}
Since $p$ covers $q$, then by Theorem 17 of \cite{Scott/Szekeres:1994}, $p$ is also a limit point of the curve $\gamma$ and thereby satisfies the requirements to be a $C^k$ scalar curvature singularity.

If $q$ is a $C^k$ non-scalar curvature singularity, then by Definition 3.4, it is a $C^k$ curvature singularity  which is not a $C^k$ scalar curvature singularity. The boundary point $p$ is also a $C^k$ curvature singularity. By the previous argument $p$ cannot be a $C^k$ scalar curvature singularity and is therefore a $C^k$ non-scalar curvature singularity.

It has thus been shown that the categories of curvature singularity, quasi-regular singularity, scalar curvature singularity and non-scalar curvature singularity all pass to the abstract boundary $\abndry{M}$.
\end{proof}

Note that $C^k$ scalar curvature singularities are clearly $C^k$ curvature singularities but the converse is not necessarily true --- i.e.\ a $C^k$ curvature singularity can be a $C^k$ non-scalar curvature singularity.
There exist space-times where the curvature components in a parallelly propagated tetrad are ill-behaved but all curvature scalar polynomials are zero \cite{Ellis/Schmidt:1977}.
One should also note the essential differences between the definitions presented above and our goal of describing physical singularities due to unbounded curvature only.
The definitions above include non-smooth behaviour other than simply the divergence of curvature components or a curvature scalar polynomial. There may be oscillatory behaviour in the curvature components or a curvature scalar polynomial and/or the derivatives of the curvature components may diverge or themselves exhibit oscillatory behaviour.

It is conceivable that fast changes in the curvature derivatives alone will not lead to the destruction of physical objects\footnote{The reader may wish to consult Ellis and Schmidt \cite{Ellis/Schmidt:1977} for a physically motivated discussion of weak curvature singularities.}.
So we must consider the question of exactly what type of curvature pathology leads to the destruction of any object nearing a singularity.
Thus, although the previous concepts of singularity given in Definitions 3.1---3.4 are both useful and physically significant, the idea of \emph{strong curvature} described in Ellis and Schmidt \cite[p.~944]{Ellis/Schmidt:1977} deals with this notion specifically.
Strong curvature has other features which make it an attractive definition of a curvature singularity to use for the proof of curvature singularity theorems.
We will now introduce the concept of strong curvature and its definition in terms of Jacobi fields.

\section{Jacobi Fields and Strong Curvature}
\begin{defn}[Jacobi Field]
If $\gamma:[0,a) \mapsto \manifold{M},\: a \in {\thereals}^+ \cup \{\infty\}$ is a geodesic with affine parameter $s$, then the smooth vector field $J:[0,a) \mapsto T\manifold{M}$  along $\gamma$ is a \emph{Jacobi field} if it satisfies the \emph{Jacobi equation}
\begin{equation*}
\vec{J}'' + \ten{R}(\vec{J},\gamma')\gamma' = 0,
\end{equation*}
where $\vec{J}'' = {\D \frac{\mathrm{D}^2 J}{\mathrm{d}s^2}} = \nabla_{\gamma'}(\nabla_{\gamma'}\vec{J})$ and $\ten{R}(\vec{X},\vec{Y})\vec{Z} = \nabla_\vec{X} \nabla_\vec{Y} \vec{Z} - \nabla_\vec{Y} \nabla_\vec{X} \vec{Z} - \nabla_{[\vec{X},\vec{Y}]}\vec{Z}$ is the curvature operator.
\end{defn}
Intuitively, a Jacobi field along $\gamma$ represents the relative displacement between $\gamma$ and nearby geodesics.
Now let $\gamma$ be a timelike geodesic with unit tangent vector $\gamma'$ and let $\vec{J}$ be a Jacobi field along $\gamma$ which is orthogonal to $\gamma'$.
Then $\gamma$ represents the path of a freely falling massive point particle and $\vec{J}$ the displacement vector between it and a second massive point particle following a nearby timelike geodesic.
Thus $\vec{J}' = \nabla_{\gamma'}\vec{J}$ represents the relative velocity and $\vec{J}''$ the relative (or tidal) acceleration of the second particle as measured by the first.
This allows us to physically interpret the Jacobi equation as linking curvature to the relative acceleration between neighbouring particles moving on geodesics.

We will now restrict our attention to Jacobi fields with particular properties.
Let $\gamma:[0,a) \mapsto \manifold{M},\: a \in {\thereals}^+ \cup \{\infty\}$ be a timelike (respectively null) geodesic with affine parameter $s$.
We define $\vec{J}_b(\gamma)$ for $b \in [0,a)$ to be the set of smooth vector fields $Z:[b,a) \mapsto T\manifold{M}$ along $\gamma$ such that
\begin{enumerate}
\item $Z(s) \in T_{\gamma(s)}\manifold{M}$,
\item $Z(b) = 0$,
\item $\D{\frac{\mathrm{D}^2 Z}{\mathrm{d}s^2} = \nabla_{\gamma'}(\nabla_{\gamma'}\vec{Z})} = \ten{R}(\gamma',\vec{Z})\gamma'$
\item $g \left( \left. \D{ \frac{\mathrm{D} Z}{\mathrm{d}s} }\right |_b, \gamma'(b) \right) = 0$
\end{enumerate}
Thus $J_b(\gamma)$ is comprised of Jacobi fields which vanish at $\gamma(b)$ and which lie in the space orthogonal to $\gamma'$ at each point $\gamma(s)$.

For a timelike geodesic $\gamma$, it is possible to choose a set of three linearly independent Jacobi fields $\{Z_1, Z_2, Z_3\}$, $Z_\alpha \in J_b(\gamma)$ ($\alpha = 1,2,3$) along $\gamma$, where for each $\alpha$, $Z_\alpha(s)$ is a spacelike vector. A spacelike 3-volume element $V(s) = Z_1(s)\wedge Z_2(s) \wedge Z_3(s)$ can then be defined at each point $\gamma(s)$.
In the case of a null geodesic $\lambda$, a set of two linearly independent Jacobi fields $\{\widehat{Z}_1, \widehat{Z}_2\}$, $\widehat{Z}_\alpha \in J_b(\lambda)$ ($\alpha = 1,2$) may be chosen along $\gamma$, where for each $\alpha$, $\widehat{Z}_\alpha(s)$ is a spacelike vector. A spacelike 2-volume element $\widehat{V}(s) = \widehat{Z}_1(s) \wedge \widehat{Z}_2(s)$ can then be defined at each point $\gamma(s)$.
We now present the strong curvature definitions of Tipler \cite{Tipler:1977} and Kr\'olak \cite{Krolak:1987}.

\begin{defn}[Strong Curvature Singularity - Tipler \cite{Tipler:1977}]
Let $\gamma:[0,a) \mapsto \manifold{M},\: a \in {\thereals}^+$ be a timelike (respectively null) geodesic with affine parameter $s$.
A \emph{Tipler strong curvature condition} is said to be satisfied along $\gamma$ if for all $b \in [0,a)$ and all linearly independent Jacobi fields $Z_1, Z_2, Z_3 \in J_b$ (respectively, $\widehat{Z}_1, \widehat{Z}_2 \in J_b$ if $\gamma$ is a null geodesic)
\begin{equation*}
 \liminf_{s \goesto a} V(s) = 0 \text{ (or, respectively, $\liminf_{s \goesto a} \widehat{V}(s) = 0$).}
\end{equation*}

\end{defn}

\begin{defn}[Strong Curvature Singularity - Kr\'olak \cite{Krolak:1987}]
Let $\gamma:[0,a) \mapsto \manifold{M},\: a \in {\thereals}^+$ be a timelike (respectively null) geodesic with affine parameter $s$.
The \emph{Kr\'olak strong curvature condition} is said to be satisfied along $\gamma$ if for all $b \in [0,a)$ and all linearly independent Jacobi fields $Z_1, Z_2, Z_3 \in J_b$ (respectively, $\widehat{Z}_1, \widehat{Z}_2 \in J_b$ if $\gamma$ is a null geodesic) there exists a $c \in [b,a)$ such that 
\begin{equation*}
 \left. \D{ \frac{\mathrm{d} V}{\mathrm{d}s} }\right |_c < 0 \;\;\;\;\;\; \left( \text{respectively,}  \left. \D{ \frac{\mathrm{d} \widehat{V}}{\mathrm{d}s} }\right |_c < 0 \right).
\end{equation*}
\end{defn}

The concept of a strong curvature singularity was first introduced by Ellis and Schmidt \cite{Ellis/Schmidt:1977}.
They loosely considered it a type of singularity at which no object could arrive intact due to the presence of unbounded tidal forces.
Tipler \cite{Tipler:1977} was the first to formalise this definition to the one above using only geometrically defined objects but with the same physical content as the earlier definition.
Essentially, Tipler's definition requires that any object have its volume crushed to zero as the singularity is approached.
The Kr\'olak \cite{Krolak:1987} definition is weaker than the Tipler version and arose out of investigations into the cosmic censorship conjecture \cite{Krolak:1986,Krolak:1992,Krolak/Rudnicki:1993}.
In this literature the Kr\'olak definition is sometimes referred to as the \emph{limiting focusing condition}.

Our primary reason for considering the use of strong curvature is the existence of the following results due to Clarke and Kr\'olak \cite{Clarke/Krolak:1985}\footnote{We recommend that the reader refer to Clarke \cite{Clarke:1993:book} instead of Clarke and Kr\'olak \cite{Clarke/Krolak:1985} for the proofs of these theorems.
The original paper is full of misleading typographic errors which have been corrected in the proofs contained in Clarke \cite{Clarke:1993:book}.}.

\begin{prop}
For both the timelike and the null cases, if the Tipler strong curvature condition is satisfied, then for some component ${R^i}_{0j0}$ of the Riemann tensor in a parallelly propagated frame the integral
\begin{equation*}
{I^i}_j(v) = \int^{v}_{0} dv' \int^{v'}_{0} dv'' |{R^i}_{0j0}(v'')|
\end{equation*}
does not converge as $v \goesto a$.
\end{prop}

\begin{prop}
If $\lambda(v)$ is a null geodesic and the Tipler strong curvature condition is satisfied then, with respect to a parallelly propagated frame, either the integral
\begin{equation*}
K(v) = \int^v _0 dv' \int ^{v'} _0 dv'' R_{00}(v'')
\end{equation*}
or the integral
\begin{equation*}
{L^m}_n (v) = \int ^v _0 dv' \int ^{v'} _0 dv'' \left( \int^{v''}_0 dv'''|{C^m}_{0n0}(v''')| \right)^2
\end{equation*}
for some $m, n$ does not converge as $v \goesto a$.
\end{prop}

Similar results dealing with the Kr\'olak 
definition were also proven by Clarke and Kr\'olak \cite{Clarke/Krolak:1985} but with one integral less in each case.

These results are significant since they pertain to parallelly propagated components of the Riemann, Ricci and Weyl curvature tensors and verify that strong curvature singularities not only imply the divergence of various parallelly propagated components of the Riemann, Ricci and Weyl tensors but also the divergence of their integrals, as defined above in Proposition 4.4 and Proposition 4.5, along geodesics approaching them.
In addition, Clarke and Kr\'olak \cite{Clarke/Krolak:1985} proved some sufficiency conditions on integrals of the Ricci and Weyl curvature tensor components which ensure the existence of Tipler and Kr\'olak-type strong curvature singularities.

In summary, strong curvature has three salient properties for application to the curvature singularity problem:

\begin{enumerate}

\item Strong curvature is a physically intuitive and mathematically precise way to define the presence of infinite tidal forces.

\item Clarke and Kr\'olak established the equivalence between the various notions of strong curvature and divergences in the parallelly propagated components of the Riemann, Ricci and Weyl tensors and in integrals of these curvature components along the geodesic.

\item The form of the strong curvature condition is similar to the trapping condition of Theorem \ref{Hawking.thm} (given below) by Hawking \cite[Theorem 3,~p.~271]{Hawking/Ellis:1973:book}.
\end{enumerate}

It is this final point which we will proceed to examine in more detail.
Previous arguments have made it clear that the production of a curvature singularity theorem will require the connection of three key concepts:
\begin{enumerate}
\item Essential Singularities
\item Curve Incompleteness
\item Unbounded Curvature
\end{enumerate}

Concept (1) is important since a singularity predicted by any resulting theorem should not be removable simply by re-embedding the space-time.
The recent Ashley and Scott result \cite[Theorem~4.12]{Ashley:phdthesis} stated below as Theorem \ref{Ashley&Scott.thm} provides the necessary, and previously unknown, link between (1) and (2).
Concept (2) is also significant since, intuitively, realistic singularities all cause the incompleteness of physically relevant curve families.

\begin{thm}
\label{Ashley&Scott.thm}
Let \spacetime{\manifold{M}}{g} be a strongly causal, $C^l$ maximally extended, $C^k$ space-time $(1 \leqslant l \leqslant k)$ and \curvefamily{C} be the family of affinely parametrised causal geodesics in \spacetime{\manifold{M}}{g}. Then the abstract boundary \curvefamily{B}$($\manifold{M}$)$ of \manifold{M} contains a $C^l$ essential singularity iff there is an incomplete causal geodesic in \spacetime{\manifold{M}}{g}.
\end{thm}

It has proven difficult to establish theorems on divergent curvature directly from the types of conditions typically employed in the geodesic incompleteness theorems.
A link between (2) and (3), however, is provided using strong curvature and the following theorem by Hawking \cite[Theorem~3,~p.~271]{Hawking/Ellis:1973:book}, \cite{Hawking:1967}.

\begin{thm}
\label{Hawking.thm}
    If the space-time \spacetime{\manifold{M}}{g} satisfies the following conditions:
    \begin{enumerate}
      \item $R_{ab}K^{a}K^{b}\geqslant 0$ for all non-spacelike vectors $K$;
      \item the strong causality condition holds on \spacetime{\manifold{M}}{g};
      \item there is some past-directed unit timelike vector $W$ at a 
      point $p$ and a positive constant $b$ such that if $V$ is the unit 
      tangent vector to the past-directed timelike geodesics through $p$, 
      then on each such geodesic the expansion $\theta \equiv 
      {V^{a}}_{;a}$ of 
      these geodesics becomes less than $-3c/b$ within a distance $b/c$ from 
      $p$, where $c \equiv -W^{a}V_{a}$,
      \end{enumerate}
then there is a past incomplete causal geodesic through $p$.
\end{thm}

The result which follows comes directly from Theorem \ref{Hawking.thm} and the proof of Theorem \ref{Ashley&Scott.thm}. Within the context of the singularity theorems, it provides the desired link between curve incompleteness and the existence of essential singularities. It says that if there is an incomplete causal geodesic $\gamma$ in \manifold{M} predicted by the Hawking theorem, then $\gamma$ corresponds to an $a$-boundary essential singularity of \spacetime{\manifold{M}}{g}. In accordance with the conditions imposed for the Penrose-Hawking singularity theorems as well as Theorem \ref{Ashley&Scott.thm}, it will be assumed that the space-time \spacetime{\manifold{M}}{g} is maximally extended.

\begin{cor}
\label{Ashley&Scott.cor}
Suppose that \spacetime{\manifold{M}}{g} is a maximally extended space-time which satisfies the conditions of the Hawking theorem.
Let $\gamma$ be a past incomplete causal geodesic through $p \in \manifold{M}$ predicted by that theorem. Then there exists a $C^\infty$ embedding $\phi:  \manifold{M} \mapsto \manifold{\widehat{M}}$ (where  \manifold{\widehat{M}} is a 4-dimensional manifold) and a boundary point $q \in \envbndry{\phi}{M} \subset \manifold{\widehat{M}}$ such that $q$ is a limit point of $\gamma$ and is an $a$-boundary essential singularity.
\label{curvature.cor}
\end{cor}

This result is important since it provides the connection between the focusing condition on the expansion of the timelike geodesic congruence through $p$ (condition 3 of Theorem \ref{Hawking.thm}) and the existence of an incomplete causal geodesic which corresponds to an essential singularity.
All that would now be required to obtain a curvature singularity theorem is to link this condition with our chosen concept of a strong curvature singularity.
Since it is possible to determine the expansion $\theta$ of a congruence of geodesics in terms of the volume element  $V(s)$  derived from the Jacobi fields, namely,
\begin{equation*}
\theta = \frac{1}{V(s)}\frac{\mathrm{d}V(s)}{\mathrm{d}s}\;,
\end{equation*}
we are able to rewrite the strong curvature definition of Kr\'olak in terms of the expansion.
This is a profitable exercise since the expansion scalar obeys the Raychaudhuri equation.
Specifically, for the case of $\hat{\theta}$ being the expansion scalar for a congruence of null geodesics, the Raychaudhuri equation takes the form \cite[p.~88]{Hawking/Ellis:1973:book}
\begin{equation*}
\frac{\mathrm{d}\hat{\theta}}{\mathrm{d}s} = -R_{ab}K^a K^b - 2\hat{\sigma}^2 - \frac{1}{2}\hat{\theta}^2\;,
\end{equation*}
where $\hat{\sigma}(s)$ is the shear of the congruence and $K^a (s)$ is the tangent vector of the null geodesic along which $\hat{\theta}$ is being determined.
If we define a length scale $\hat{x}^2 = \hat{V}(s)$ where $\hat{V}(s)$ is the spacelike 2-volume element then
\begin{equation*}
\hat{\theta} = \frac{1}{\hat{V}(s)}\frac{\mathrm{d}\hat{V}(s)}{\mathrm{d}s} = \frac {2}{\hat{x}}\frac{\mathrm{d}\hat{x}}{\mathrm{d}s}\;,
\end{equation*}
and the Raychaudhuri equation becomes
\begin{equation}
\frac{\mathrm{d}^2\hat{x}}{\mathrm{d}s^2}=-\frac{1}{2}[R_{ab}K^a K^b + 2 \hat{\sigma}^2] \hat{x}\;.
\label{ray.eqn}
\end{equation}

Pursuing the details of this propagation equation for the two definitions of the strong curvature condition we obtain the following lemma and definition.
\begin{lem}[Tipler Strong Curvature Condition {\cite[p.~164]{Clarke:1993:book}}]
Let $\lambda:[0,a) \mapsto \manifold{M},\: a \in {\thereals}^+$ be a null geodesic with affine parameter $s$.
The Tipler strong curvature condition implies that for all $b \in [0,a)$ and all solutions $\hat{x}(s)$ of Equation \eqref{ray.eqn} with initial conditions $\hat{x}(b)=0$, $\liminf_{s \goesto a} |\hat{x}(s)| = 0$.
\end{lem}
\begin{defn}[Kr\'olak Strong Curvature Condition {\cite{Krolak:1992}}]
\label{KrolakSCC.defn}
Let $\lambda:[0,a) \mapsto \manifold{M},\: a \in {\thereals}^+$ be a future-inextendible null geodesic with affine parameter $s$.
It will be said that $\lambda$ terminates in a \emph{strong curvature singularity} in the future if, for all points $p$ on $\lambda$ and for all initial value conditions for the Raychaudhuri equation and the shear equation at $p$, there exists a point $q$ on $\lambda$ to the future of $p$ such that $\hat{\theta}  = [2/\hat{x}(s)][\mathrm{d}\hat{x}(s)/\mathrm{d}s]$ is negative.
\end{defn}
It is the version of Definition \ref{KrolakSCC.defn} relating to a past-inextendible null geodesic which looks especially promising for application to Corollary \ref{Ashley&Scott.cor}.
The results in the Clarke and Kr\'olak paper \cite{Clarke/Krolak:1985} would then link this to the presence of parallelly propagated curvature components which are unbounded.

\section{Future Directions in Proving Curvature Singularity Theorems}
This paper has outlined a program within which curvature singularity theorems might be proven.
In order to complete this program, a specific definition of strong curvature must be chosen.
Physical intuition would lead one to choose the Tipler condition, however it may be easier in practice to use the Kr\'olak condition.
Later extensions which include the Tipler condition should then be relatively straightforward.

In conclusion, it appears that this program of research, incorporating the use of the abstract boundary construction, offers some exciting opportunities for proving curvature singularity theorems.
Given the importance of the long-standing investigation into singularities, geodesic incompleteness and curvature, such a result would form the critical step in determining the physical content of singularity production in General Relativity.

\bibliographystyle{amsplain}
\bibliography{ashleyscottreferences}

\providecommand{\bysame}{\leavevmode\hbox to3em{\hrulefill}\thinspace}
\providecommand{\MR}{\relax\ifhmode\unskip\space\fi MR }
\providecommand{\MRhref}[2]{%
  \href{http://www.ams.org/mathscinet-getitem?mr=#1}{#2}
}
\providecommand{\href}[2]{#2}
\begin{thebibliography}{10}

\bibitem{Ashley:phdthesis}
Michael John Siew~Lueng Ashley, \emph{Singularity theorems and the abstract
  boundary construction}, Ph.D. thesis, The Australian National University,
  2002.

\bibitem{Belinskii/Khalatnikov/Lifshitz:1970}
V.~A. Belinskii, I.~M. Khalatnikov, and E.~M. Lifshitz, \emph{Oscillatory
  approach to a singular point in relativistic cosmology}, Adv. in Phys.
  \textbf{19} (1970), 523--573.

\bibitem{Bosshard:1979}
B.~Bosshard, \emph{On $b$-boundaries of special space-time models}, Gen. Rel.
  Grav. \textbf{10} (1979), 963--966.

\bibitem{Clarke:1975}
C.~J.~S. Clarke, \emph{The classification of singularities}, Gen. Rel. Grav.
  \textbf{6} (1975), 35--40.

\bibitem{Clarke:1993:book}
\bysame, \emph{The analysis of space-time singularities}, Cambridge University
  Press, Cambridge, 1993.

\bibitem{Clarke/Krolak:1985}
C.~J.~S. Clarke and A.~Kr{\'o}lak, \emph{Conditions for the occurrence of
  strong curvature singularities}, J. Geom. Phys. \textbf{2} (1985), 127--143.

\bibitem{Clarke/Schmidt:1977}
C.~J.~S. Clarke and B.~G. Schmidt, \emph{Singularities: The state of the art},
  Gen. Rel. Grav. \textbf{8} (1977), 129--137.

\bibitem{Ellis/Schmidt:1977}
G.~F.~R. Ellis and B.~G. Schmidt, \emph{{Singular space-times}}, Gen. Rel.
  Grav. \textbf{8} (1977), 915--953.

\bibitem{Ellis/Schmidt:1979}
\bysame, \emph{{Classification of singular space-times}}, Gen. Rel. Grav.
  \textbf{10} (1979), 989--997.

\bibitem{Geroch:1968a}
R.~Geroch, \emph{Local characterization of singularities in general
  relativity}, J. Math. Phys. \textbf{9} (1968), 450--465.

\bibitem{Geroch/Kronheimer/Penrose:1972}
R.~Geroch, E.~H. Kronheimer, and R.~Penrose, \emph{Ideal points in space-time},
  Proc. R. Soc. Lond. \textbf{A327} (1972), 545--567.

\bibitem{Hawking:1967}
S.~W. Hawking, \emph{The occurrence of singularities in cosmology. {III
  Causality and singularities}}, Proc. R. Soc. Lond. \textbf{A300} (1967),
  187--201.

\bibitem{Hawking/Ellis:1973:book}
S.~W. Hawking and G.~F.~R. Ellis, \emph{The large scale structure of
  space-time}, Cambridge University Press, 1973.

\bibitem{Krolak:1992}
A.~Kr{\'{o}}lak, \emph{Strong curvature singularities and causal simplicity},
  J. Math. Phys. \textbf{33} (1986), 701--704.

\bibitem{Krolak:1986}
\bysame, \emph{Towards the proof of the cosmic censorship hypothesis}, Class.
  Quant. Grav. \textbf{3} (1986), 267--280.

\bibitem{Krolak:1987}
\bysame, \emph{Towards the proof of the cosmic censorship hypothesis in
  cosmological space-times}, J. Math. Phys. \textbf{28} (1987), 138--141.

\bibitem{Krolak/Rudnicki:1993}
A.~Kr{\'{o}}lak and W.~Rudnicki, \emph{Singularities, trapped sets, and cosmic
  censorship in asymptotically flat space-times}, Inter. J. Theor. Phys.
  \textbf{32} (1993), 137--142.

\bibitem{Lifshitz/Khalatnikov:1963}
E.~M. Lifshitz and I.~M. Khalatnikov, \emph{Investigations in relativistic
  cosmology}, Adv. in Phys. (Phil. Mag. Suppl.) \textbf{12} (1963), 185--249.

\bibitem{Misner:1967}
C.~W. Misner, \emph{{Taub-NUT spaces as a counterexample to almost anything}},
  Relativity Theory and Astrophysics I: Relativity and Cosmology (J.~Ehlers,
  ed.), Lectures in Applied Mathematics, vol.~8, American Mathematical Society,
  1967, pp.~160--169.

\bibitem{Penrose:1965c}
R.~Penrose, \emph{Gravitational collapse and space-time singularities}, Phys.
  Rev. Lett. \textbf{14} (1965), 57--59.

\bibitem{Schmidt:1971}
B.~G. Schmidt, \emph{A new definition of singular points in general
  relativity}, Gen. Rel. Grav. \textbf{1} (1971), 269--280.

\bibitem{Scott/Szekeres:1994}
S.~M. Scott and P.~Szekeres, \emph{The abstract boundary--a new approach to
  singularities of manifolds}, J. Geom. Phys. \textbf{13} (1994), 223--253.

\bibitem{Szabados:1988}
L.~Szabados, \emph{Causal boundary for strongly causal space-times}, Class.
  Quant. Grav. \textbf{5} (1988), 121--134.

\bibitem{Tipler:1977}
F.~J. Tipler, \emph{Singularities in conformally flat spacetimes}, Phys. Lett.
  \textbf{64A} (1977), 8--10.

\end{thebibliography}

\end{document}